\documentclass[useAMS,usenatbib]{mn2e}
\usepackage{natbib}
\usepackage[usenames]{color}

\usepackage{amssymb}
\usepackage{amsmath}

\usepackage[english]{babel}
\usepackage{graphicx}
\newcommand{\mnras}{MNRAS}
\newcommand{\apj}{ApJ}
\newcommand{\aap}{A\&A}

\newcommand{\apjl}{ApJ}
\newcommand{\araa}{ARA\&A}
\newcommand{\aj}{AJ}

\def\msun{\,\rm {M}_{\odot}}

\def\simlt{\mathrel{\rlap{\lower 3pt\hbox{$\sim$}}\raise 2.0pt\hbox{$<$}}} 
\def\simgt{\mathrel{\rlap{\lower 3pt\hbox{$\sim$}} \raise 2.0pt\hbox{$>$}}} 
\def\lsim{\mathrel{\rlap{\lower 3pt\hbox{$\sim$}}\raise 2.0pt\hbox{$<$}}} 
\def\gsim{\mathrel{\rlap{\lower 3pt\hbox{$\sim$}} \raise 2.0pt\hbox{$>$}}} 
\def\lta{\mathrel{\rlap{\lower 3pt\hbox{$=$}}\raise 2.0pt\hbox{$<$}}} 
\def\gta{\mathrel{\rlap{\lower 3pt\hbox{$=$}} \raise 2.0pt\hbox{$>$}}} 
 
\newcommand{\beq}{\begin{equation}}
\newcommand{\eeq}{\end{equation}}
\newcommand{\h}{{\rm h}}

\begin{document} 
 
\title{High redshift formation and evolution of central massive
  objects I: model description } \author[Devecchi, et al.]{B.~
  Devecchi$^1$, M. Volonteri$^2$, M. Colpi$^3$, F. Haardt$^1$\\
  $^1$ Dipartimento di Fisica \& Matematica, Universit\`a dell'Insubria,
  Via Valleggio 11, 22100 Como, Italy\\ $^2$ Dept. of Astronomy,
  University of Michigan, Ann Arbor, MI 48109, USA\\ $^3$ Dipartimento di
  Fisica G.~Occhialini, Universit\`a degli Studi di Milano Bicocca,
  Piazza della Scienza 3, 20126 Milano, Italy}

\maketitle \vspace {7cm}
 
\begin{abstract} 
Galactic nuclei host central massive objects either in the form of
supermassive black holes or nuclear stellar clusters. Recent
investigations have shown that both components co-exist in at least a
few galaxies. In this paper we explore the possibility of a connection
between nuclear star clusters and black holes that establishes at the
moment of their formation. We here model the evolution of high
redshift discs, hosted in dark matter halos with virial temperatures
$> 10^4$ K, whose gas has been polluted with metals just above the
critical metallicity for fragmentation. A nuclear cluster forms as a
result of a central starburst from gas inflowing from the unstable
disc.  The nuclear stellar cluster provides a suitable environment for
the formation of a black hole seed, ensuing from runaway collisions
among the most massive stars.  Typical masses for the nuclear stellar
clusters at the time of black hole formation ($z\sim 10$) are in
the range $10^4-10^6\,\msun$ and have half mass radii $\lsim 0.5$
pc. The black holes forming in these dense, high redshift clusters can
have masses in the range $\sim 300-2000 \msun$.

\end{abstract}


\section{Introduction} 

Central massive objects are known to reside in the centers of a large
fraction of galaxies either in the form of supermassive black holes
(BHs) or nuclear stellar clusters (NCs). Both BHs and NCs have masses
that correlate with those of their hosts (Ferrarese et al. 2006; 
Wenher et al. 2006), suggesting a symbiotic evolution between these
nuclear components and the larger structure of their galaxy hosts.

NCs, with masses, $M_{\rm NC}$, ranging between $10^5-10^7\,\msun$
(Walcher et al. 2006; C{\^o}t{\'e} et al.(2006); Bokert et al. 2004;
Balcells et al. 2007) appear to be more common in galaxies with
stellar masses below $\sim 10^{11}\,\msun$, while BHs ($M_{\rm
  BH}>10^6\,\msun$) preferentially inhabit galaxies more massive than
$10^{10}\,\msun$ (Ferrarese et al. 2006; Wenher et al. 2006). Despite
this apparent dichotomy, a smooth transition exists between NCs and
BHs.  Some galaxies do host both a NC and an embedded BH (Seth et
al. 2008; Graham et al. 2009) and observational estimates seems to
indicate mass ratios $M_{\rm BH}/M_{\rm NC}\sim\,10^{-4}-10^{-1}$
(Seth et al. 2008). Interestingly, BHs coexisting with NCs have masses
that covers the intermediate range ($10^3-10^6\,\msun$) between
stellar and supermassive BHs.

Two main scenarios have been proposed to explain the origin of NCs.
In the first, NCs are the end product of multiple mergers of globular
clusters following their sinking by dynamical friction into the
central part of the host galaxy (Tremaine et al. 1975;
Capuzzo-Dolcetta \& Miocchi 2008; Lotz et al. 2001). In the second,
NCs form {\it in situ} after inflow of gas from the outer part of the
host (Milosavljevic 2001): this latter possibility seems more
consistent with stellar population studies in NCs (C{\^o}t{\'e} et
al.(2006)).

Whereas models of NCs in a cosmological context are still lacking,
models of BH seed formation have been developed in the $\Lambda$CDM
cosmogony. How BH seeds form and how they evolve into the supermassive
variety has been the subject of many studies (Begelman \& Rees 1986).
A natural path to BH formation relies on the first generation of stars
(Pop~III) born in metal free environments (Madau \& Rees 2001; 
Volonteri et al. 2003). These stars are expected to be more massive
than their current counterparts (Abel et al. 2000;  Omukai \& Palla
2003; Freese et al. 2008; Iocco et al. 2008).  PopIII stars that form
above a threshold mass of $\simeq 260 \,M_{\odot}$ are believed to
directly implode leaving a seed BH of similar mass (Heger et
al. 2003).  Alternative routes to BH seed formation typically exploit
gas-dynamical processes in metal-free galaxies (direct formation
models) (Haehnelt \& Rees 1993; Loeb \& Rasio 1994; Eisenstein \& Loeb
1995; Bromm \& Loeb 2003; Koushiappas et al. 2004; Begelman, Volonteri
\& Rees 2006; Lodato \& Natarajan 2006). Gravitational instabilities
can indeed lead to a vigorous gas inflow into the very central region
of protogalactic discs, supplying the necessary matter for the
formation of BH seeds.

The above scenarios have a common fundamental ingredient, i.e., the
gas hosted in the progenitor halo has not been polluted by
metals. Metal pollution marks the end of the formation of
PopIII stars and of their seed black holes as fragmentation and
formation of low mass stars start as soon as gas is polluted above a
critical metallicity threshold, $Z_{\rm crit}$ (Schneider et al.
2006; Bromm et al. 1999, 2002; Omukai et al. 2008; Clark et al. 2008,
Santoro \& Shull 2006).  
PopIII stars with masses between 140 and 260 $\msun$ are believed 
to be responsible for starting metal pollution, as they do not
leave a relic black hole but explode releasing all metals in the
surrounding medium. Star formation is also an impediment for
strong gaseous inflows in the center of massive halos as it limits the
available mass supply. This does not necessary means the end of the seed
formation epoch as a new channel for BH formation opens (Omukai et
al. 2008; Devecchi \& Volonteri 2009, hereafter DV09).

DV09 showed that the first episode of low-mass, PopII star formation can provide
the conditions for the formation of NCs and BH seeds arising in the
most compact of these clusters. Gravitationally unstable protogalactic
discs whose gas is only mildly polluted by metals, can still produce 
strong gas inflows into their central regions.
Stars start to form in the central few pc where the gaseous mass has
been accumulated and the densities are high enough for star formation
to set in.  Clusters formed in this way are crowded places.  Star-star
collisions in their core can proceed in a runaway fashion and a very
massive star of $\sim 1000\, \msun$ can build up before the first
supernova explodes (Freitag et al. 2006b,a; Portegies Zwart \& 
McMillan 2002; Gurkan et al. 2004).  At low metallicities, the final
fate of this very massive star is to collape into a BH with
mass similar to its progenitor -- as mass loss at low metallicity is negligible.

One of the factors that regulate the competition, at a given redshift,
between different formation scenarios is the metal enrichment history.
Both observational and theoretical works have been performed in order
to investigate how metals are distributed as cosmic structures form
and evolve (Scannapieco et al. 2003; Tornatore et al. 2006; Savaglio
2006; Savaglio et al. 2005; Prochaska et al. 2003; Prochaska et
al. 2007, Kulkarni et al. 2005; Li 2007). These studies have shown
that metal enrichment proceeds in a very inhomogeneous fashion.
This opens the possibility that different modes of BH seed formation
co-exist in different regions of the Universe at the same redshift,
instead of being mutually exclusive.  As the mean metallicity
increases with cosmic time, growing bubbles of polluted gas where the
first NCs and their BH seeds form, co-exist with relatively pristine
regions where PopIII star formation can still be feasible.

This is the first of a series of papers in which we develop a model
aimed at tracing self-consistently metal enrichment, following the
hierarchical build-up of halos. We consider metal enrichment and
radiative feedback from the first stars together with feedback of
successive generations of stars.  Our study is applied in the context
of BHs and NCs formation in order to determine: (i) if one of the
mechanisms outlined above is dominant in the overall seed
formation scenario; (ii) if a clear transition in redshift exists
between different formation channels or if they co-exist in the same
cosmic epoch; (iii) the characteristic redshift at which each process
reaches its maximum efficiency; (iv) when the first NCs are expected
to form.

In order to address the above questions, we develop in this paper the
semi-analytical scheme, aimed at describing the evolution of a single
halo in which a central massive object forms. We adopt a $\Lambda$CDM
cosmology with $\Omega_{\rm b} $ = 0.041, $\Omega_{\rm m}$ = 0.258,
$\Omega_{\Lambda}$ = 0.742, $h$ = 0.742 and $n_{\rm s}$ = 0.963 as
given by five-year WMAP data (Dunkley et al. 2009).

The outline of the paper is as follows. In Section 2 we descibe our
recipes for the formation and evolution of a gaseous disc, hosted in a
high redshift dark matter halo. In Section 3 we set our model for the
formation of NCs resulting from dynamically unstable discs and of BH
seeds ensuing from runaway collapse of stars in NCs.  In Section 4 we
illustrate the results and compare the results of DV09 with those
found with our new technique. Finally, in Section 5 we briefly
summarise our model.

In the subsequent papers we will apply our semi-analytical code to
merger tree histories extracted from the code PINOCCHIO (Monaco et
al. 2002a,b). This will allow us to trace halo evolution, keeping
track also of their spatial positions, a condition necessary in order
to understand the relevance of radiative and SNae feedback.

\section{Disc  formation} 

We here consider a dark matter halo of mass $M_\h$, virialising at
redshift $z_{\rm vir}$, assuming that it follows the density profile
of an isothermal sphere. Virial radius $R_{\rm vir}$, circular
velocity $V_\h$ and virial temperature $T_{\rm vir}$ are inferred from
the top hat collapse model (see for example Barkana \& Loeb 2001).

Gas falling into the potential well created by the dark matter is
heated by adiabatic compression and shocks, attaining a mean
temperature equal to $T_{\rm vir}$.  We assign an initial gas mass
$(\Omega_{\rm b}/\Omega_{\rm m})\,M_{\rm h}$ and assume that gas
follows an isothermal density profile $\rho_{\rm gas}({\rm r})$.

It has been proposed that a critical metallicity $Z_{\rm crit}$ exists
above which star formation in the low mass mode starts. We here focus
only on those halos whose gas has been enriched above $Z_{\rm crit}$.
In the following we describe the details of our treatment for the
evolution of halo gaseous component, star formation and BH seed
formation.

\subsection{Gas cooling}

We assume the standard model for disc formation (White \& Rees (1978),
Cole et al. (2000), Mo, Mao \& White 1998).  To this purpose for a
given $T_{\rm vir}$ and metallicity $Z$ of the gas, the cooling
timescale $\tau_{\rm cool}$ can be computed as

\begin{equation}
\tau_{\rm cool}=\frac{3}{2}\frac{\mu m_{\rm H}}{\rho_{\rm
    gas}(r)}\frac{k_{\rm B}T_{\rm vir}}{\Lambda (T_{\rm vir},Z)},
\end{equation}
\noindent
where $\Lambda (T_{\rm vir},Z)$ is the cooling function. We adopt
$\Lambda (T_{\rm vir},Z)$ tabulated in Sutherland \& Dopita
(1993). The cooling radius $r_{\rm cool}(t)$ at time $t$ is defined by
imposing $\tau_{\rm cool}=t$.  Matter inside $r_{\rm cool}$ has time
enough to cool down from its initial temperature.

For the disc to form, gas has to both cool down and collapse. The
timescale for disc formation will therefore be the longest between the
cooling and the free-fall timescales. In analogy with the computation
of $r_{\rm cool} (t)$ we can define a free-fall radius $r_{\rm ff}(t)$
that sets the radius inside which the gas has sufficient time to
free-fall into the central structure.  During the initial phase of the
collapse a mass of cold gas, $M_{\rm cold},$ develops at a rate

\begin{equation}
  \dot{M}_{\rm cold}=4\pi\rho_{\rm gas}(r_{\rm min}(t))r^2_{\rm
    min}(t)\dot{r}_{\rm min},
\label{eq:mcoolrate}
\end{equation}
\noindent
where $r_{\rm min}={\rm min}[r_{\rm cool},r_{\rm ff}]$.

The cold gas is assumed to condense into a rotationally supported disc
whose angular momentum $J_{\rm d}$ is a fraction $j_{\rm d}$ of the
angular momentum $J_\h$ of the halo. If the specific angular momentum
of baryons is conserved during collapse, $j_{\rm d}=M_{\rm
  cold}/M_\h$.
  
The surface density profile of the disc follows a Mestel, isothermal
profile: $\Sigma(R)=\Sigma_0(R_0/R)$ where scale parameters $\Sigma_0$
and $R_0$ are calculated assuming mass and angular momentum
conservation (see DV09)\footnote{Note that the evolution of the
  collapsing structure and the initial assembly of the disc is
  described in a way similar to DV09.  The main difference is that the
  assembly of the disc is here followed solving esplicitly for the
  time evolution of the cold gas, instead of assuming that a fixed
  fraction of the halo gas cools down instantaneously. }.

In the following Sections we describe how scale parameters of the disc
vary with cosmic time since (i) the disc mass keeps increasing due to
the infall of fresh cold gas accreting from the halo (as described by
equation \ref{eq:mcoolrate}) and (ii) the gas content changes due to
gravitational instabilities that triggers episodes of inflow and star
formation (outlined in Section \ref{sec:gi}).

\subsection{Gravitational instabilities}\label{sec:gi}

The stability of the disc is described in terms of the Toomre
parameter $Q$ (Toomre 1964). As the surface density increases since
more cold gaseous mass is added to the disc, $Q$ decreases and
eventually drops below the critical value for disc stability $Q_{\rm
  c}$.  We here assume $Q_{\rm c}=2$ (see DV09, Lodato \& Natarajan
2007, and Volonteri et al. 2008 for a discussion on $Q_{\rm
  c}$). At this point the disc develops bar-like structures that can
lead to a redistribution of mass and angular momentum.

Inflow of gas driven by gravitational instabilities can be described
in terms of an effective viscosity $\nu$. The inflow rate is
given by (Lin \& Pringle 1987, see also Lodato 2008):

\begin{equation}\label{eq:dotinf}
\dot{M}_{\rm grav}=2\pi\nu\Sigma(R)|\frac{{\rm
      d}{\rm ln}\Omega (R)}{{\rm d} {\rm ln} R}|,
\end{equation}
\noindent
where $\Omega$ is the angular velocity and 

\begin{equation}
\nu=\eta\left(\frac{Q^2_c}{Q^2}-1\right)\frac{c^3_s}{\pi G\Sigma(R)}.
\end{equation}
\noindent
We set in our reference model $\eta=0.3$ (Gammie 2001) but we also
compare our result with simulations run with $\eta=1$ (high efficiency
inflow) and $\eta=0.1$ (low efficiency inflow). The sound speed
$c_{\rm s}$ corresponds to a gas with temperature of 8000 K\footnote{
  In this paper we consider only halos illuminated by a flux in the
  Lyman-Werner (LW) band (11.2-13.6 eV). LW photons can
  photo-dissociate $H_2$ molecules, thus suppressing cooling below
  8000 K, as dictated by Ly$\alpha$ cooling. Gas could in principle
  cool down to much lower temperature in presence of $H_2$. Less
  massive, colder discs form, leading to a much lower inflow. This
  prevents the formation of BH seeds unless higher inflow efficiencies
  are considered.  We will consider this more general case in our next
  paper, where LW flux ensuing from star formation history in a
  cosmological context will be calculated.}.

The inflow rate that channels mass in the central region of the disc
is then:

\begin{eqnarray}
\dot{M}_{\rm grav}&=&\frac{2\eta
  c^3_s}{G}\left[\left(\frac{Q_{\rm c}}{Q}\right)^2-1\right]=\nonumber\\
&=&\frac{\eta c_{\rm s}}{G
  V^2_\h}\left[\pi^2G^2Q^2_c\Sigma^2_0R^2_0-2c^2_sV^2_\h\right].
\label{eq:inflow}
\end{eqnarray}
\noindent

Following DV09 we assume that the inner surface density profile
changes its scaling with radius as $\propto R^{-5/3}$, according to
simulations by Mineshige \& Umemura (1999). The transition radius,
$R_{\rm tr}$, between the two profiles is given by:

\begin{equation}\label{eq:rtr}
R_{\rm tr}=\frac{M_{\rm inf}}{4\pi \Sigma_0 R_0},
\end{equation}
\noindent
where $M_{\rm inf}$ is the net mass inflowing from the outer disc
($R>R_{\rm tr}$, see DV09 and equation \ref{eq:inflow2}).

For a given set of $M_\h$, $T_{\rm gas}$ and $Q_{\rm c}$, a maximum
angular momentum exists below which a disc becomes unstable $\lambda
<1/8\Omega_b/\Omega_mQ_c\sqrt{T_{\rm vir}/T_{\rm gas}} $ where $T_{\rm
  gas}$ is temperature of the cold gaseous disc.  Discs with higher
rotation are more stable agains fragmentation as the amount of mass
that needs to be removed from the Mestel disc in order to reach
$Q=Q_{\rm c}$ is lower. The inflow mass decreases for halos with large
spin parameters.

\subsection{Star formation}

If the disc grows strongly unstable, star formation sets in, consuming
part of the gas that would otherwise flow into the central region.
Further evolution of $M_{\rm inf}$ is then regulated by the
competition between star formation in the outer Mestel disc and funneling of
gas in its center.

An unstable disc starts fragmenting into bound clumps once the
gravitationally induced stress exceeds a critical value corresponding
to $\dot{M}_{\rm crit}=0.06\,c^3_s/G$ (Lodato \& Natarajan 2007).  We
allow fragmentation in the Mestel disc only in those regions where
the coolign rate $\Lambda$ exceeds the adiabatic heating rate (see the
discussion in DV09). This condition determines the location of a
radius $R_{\rm SF}$ where stars are allowed to form.

We denote  the time lag between cloud formation and
collapse to form stars as $t_{\rm lag}$. This is basically the lifetime of
molecular clouds; in our model it corresponds to the delay between the
onset of instability (that we assume to be the driver leading to the
formation of the clouds) and the start of the star formation event.
Observational estimates for lag times range from 1 to 7 Myrs (Klessen
et al. 2009). We assume $t_{\rm lag}=3.5$ Myr. We have checked that
our results are not strongly affected by the assumed value of $t_{\rm
  lag}$: final black hole seed masses change by a factor 2-3 for
$t_{\rm lag}$ ranging between 1 and 7 Myrs.

We assume a star formation rate surface density $\Sigma_{\rm SFR}$
that follows the empirical Schmidt-Kennicutt law (Kennicutt 1998):

\begin{equation}
\Sigma_{\rm SFR}=2.5\cdot 10^{-4} \left(\frac{\Sigma(R)}{1\msun
  {\rm pc}^{-2}}\right)^{1.4} \msun {\rm yr}^{-1} {\rm kpc}^{-2},
\end{equation}
\noindent
and calculate the star formation rate $\dot{M}_{*,\rm d}$ by integrating
over the region of the disc where stars form. 
We thus integrate $\Sigma_{\rm SFR}$ between $R_{\rm tr}$ and $R_{\rm
  SF}$ to infer a star formation rate $\dot{M}_{*,\rm d}$ in the outer
disc:

\begin{equation}
\dot{M}_{*,\rm d}=5\pi\times
10^{-4}\left[\Sigma_0R_0\right]^{7/5}(R^{3/5}_{\rm SF}-R^{3/5}_{\rm
  tr}).
\label{eq:sfrate}
\end{equation}

Star formation decreases the amount of gas that can be funnelled in
the inner disc. For this reason and as long as $\dot{M}_{\rm
  grav}>\dot{M}_{\rm crit}$ we assume a net inflow rate $\dot{M}_{\rm
  inf}$ equal to

\begin{equation}\label{eq:inflow2}
\dot{M}_{\rm inf}=\dot{M}_{\rm grav}-\dot{M}_{*,\rm d}.
\end{equation}

\subsection{Supernova feedback}

Approximately 3 Myrs after an episode of star formation, the most
massive stars explode as SNae. Metals processed in their centers are
released into the surrounding gas, increasing the metal content.
Metal yields and energy production depend on the initial mass function
(IMF) of the stars. We here assume a Salpeter IMF with minimum ($M_{\rm l}$)
and maximum ($M_{\rm u}$) mass of 0.1 and 100 $\msun$ respectively
normalized to 1 $\msun$.

We correlate the rate of metal production with the star formation rate
in the disc assuming that the total  mass in metals scales as:

\begin{equation}
\dot{M}_{\rm met}=Y_{\rm met} \nu_{\rm SN} \dot{M}_*,
\end{equation}
\noindent
where $Y_{\rm met}$ is the total IMF-averaged metal yield and
$\nu_{\rm SN}$ is the fractional number of SNae exploding after the
formation of a mass $M_*$ in stars.  We assume that SN progenitors are
stars with masses between 10-50 $\msun$ (Scannapieco et
al. 2003)\footnote{We here follow Tsujimoto et al 1995 and assume that
  stars with masses greater than 50 $\msun$ collapse into a BH without
  ejecting heavy metals.}. With our prescriptions $Y_{\rm met}=1.6
\msun$ and $\nu_{\rm SN} =0.00484 $.

SN explosions in high redshift halos can be extremely destructive. SN
driven bubbles can push away part of the gas in the halo, eventually
fully depriving the host of its gas. The efficiency of gas depletion
depends on the depth of the potential well of the host and on the
energy released during the explosion.

Using energy conservation we compute the amount of mass that is
removed from the disc plus halo as a result of SN esplosions, assuming
that each single explosion releases an energy $E_{\rm SN}=10^{51}$ erg
(Woosley \& Weaver 1995). The resulting outflow rate $\dot{M}_{\rm
  sh}$ is

\begin{equation}
\dot{M}_{\rm sh}=\frac{f_{w}\nu_{\rm SN} E_{\rm SN}\dot{M}_{*}
}{2\left(1+\Omega_{\rm b}/\Omega_{\rm m}\right)GM_\h/R_{\rm
    vir}+\frac{1}{2}v^2_{\rm sh}},
\end{equation}
\noindent
where $f_{w}$ is the fraction of energy channelled into the outflow,
$v_{\rm sh}$ is the velocity of the outflow at the virial radius. We
refer the reader to Appendix A for details of our recipes for gas depletion.
Lastly, we define the retention fraction coefficient $f_{\rm
  ret}\equiv (M_{\rm gas}-M_{\rm sh})/M_{\rm gas}$.  After the
explosion we assume that a fraction $f_{\rm ret}$ of the metals yields
produced remains in the host.

\subsection{Disc evolution}

The evolution of an unstable disc is followed tracing the time
dependence of $M_{\rm cold}$, $M_{\rm inf}$, $M_{\rm *,d}$ and $M_{\rm
  sh}$, integrating equations 2, 8, 9 and 11 assuming that initially
all the gaseous mass is hot ($T=T_{\rm vir}$).

Changes in these quantities over cosmic time correspond to changes in
the scale parameters of the Mestel disc ($\Sigma_0$, $R_0$) and
$R_{\rm tr}$. Angular momentum conservation and total mass
conservation in the system are imposed to infer the following
equations linking the current values of $\Sigma_0(t)$ and $R_0(t)$ to
their value at the moment in which the disc reaches the instability
regime:

\begin{equation}
\Sigma_0(t)=\Sigma_0(0)\left[1-\frac{m_{\rm a}(t)}{m_{\rm d}(t)}\right]^{3},
\end{equation}

\begin{equation}
R_0(t)=R_0(0)\left[1-\frac{m_{\rm a}(t)}{m_{\rm d}(t)}\right]^{-1}
\end{equation}
\noindent
where $m_{\rm d}(t) = M_{\rm cool}(t)/M_\h$ and $m_{\rm a}(t)$ is the
total fractional gaseous mass removed from the outer disc, either via
star formation, inflow and outflows from SNae.  The disc alternates
episodes of instability to states of quiescence: as $Q$ increases
above $Q_{\rm c}$, the instability is quenched. Disc parameters evolve
regulated by the amount of new mass that is added by accretion from
the halo.

We halt the simulations as soon as the hot halo gas is consumed and
the disc is in a stable state.  We proceed on studying in the next
section the evolution of the gas that has been funnelled at the center
of the disc where we expect that a NC and/or a BH seed form.

\section{Central massive object formation}

In the previous Section we focused on the evolution of Toomre unstable
discs. Here we study the formation of a NC from a mass reservoir
$M_{\rm inf}$, funnelled into $R_{\rm tr}$ after $t_{\rm lag}$.

Star formation in the central part of the disc can proceed with a
higher efficiency with respect to the outer disc as higher pressure
and density are expected there (Elmegreen \& Efremov 1997, Li et
al. 2007). We assume that a fraction $\epsilon_{\rm SF}$ of $M_{\rm
  inf}$ is converted into stars and we here adopt $\epsilon_{\rm
  SF}=1$ (Li et al. 2007). The mass and radius of the cluster are
$M_{\rm cl}=\epsilon_{\rm SF}M_{\rm inf}$ and $R_{\rm cl}={\rm
  min}(R_{\rm SF},R_{\rm tr})$.

NCs formed in this way are very compact and unstable
objects. Numerical studies of stellar clusters have shown that
gravitational encounters between stars can drive the clusters towards
core collapse, leading to a contraction of the core where physical
collisions between stars are frequent\footnote{Note that these studies
  concentrate on halo stellar clusters. NC dynamics can be alterated
  with respect to their halo conterparts by (i) the presence of the
  external stellar environment and (ii) continuos gas accretion. These
  two effects work on opposite directions: cluster contraction can be
  reverted as stars of the more rarefied outer-system are trapped into
  the central clusters. Inflow of gas can instead drive an
  acceleration of core collapse as it deepens the potential well. The
  final outcome resulting from the competition between them is not
  easy to infer and it is behind the aim of this paper. We will
  address this issue in a follow-up paper.}. Core collapse is driven
by dynamical friction of the more massive stars. Star-star collisions
between massive stars in the cluster core can happen in a runaway
fashion, leading to the build up of a very massive star (VMS), whose
mass can be as high as 1000 $\msun$ (Portegies Zwart et al. 1999,2004;
Freitag et al. 2006a,b; Gurkan et al. 2004).

After $\sim 3 $ Myrs the first SNae explode. For the systems we are
interested in, this happens before the inflow process ends. We assume
that a NC is able to produce a BH seed only if it undergoes core
collapse before the first SN explosion.  We compute the mass of the
nuclear star cluster, $M_{\rm cl,0},$ 3 Myr after the onset of star
formation, and from these we infer the core collapse timescale,
$t_{\rm cc}$. For those systems with $t_{\rm cc}\leq 3$ Myr, we
calculate the mass for the very massive star $M_{\rm VMS}$ as in DV09:

\begin{equation}
M_{\rm VMS}=m_*+4\times 10^{-3}M_{\rm
  cl}f_c\ln{\lambda_C}\ln{\left(\frac{3{\rm Myr}}{t_{\rm
      cc}}\right)},
\end{equation}
\noindent
where $f_{\rm c}=0.2$ and $\ln{\lambda_C}=\log(0.1 M_{\rm cl,0}/\langle
m\rangle)$ (Portegies Zwart \& McMillan (2002)).  

At low metallicities, stars more massive than 260 $\msun$ are expected
to leave a remnant BH, that retains most of its progenitor mass.  Each
time $M_{\rm VMS}$ is greater than 260 $\msun$ we assume a BH seed is
formed with $M_{\rm BH}=M_{\rm VMS}$.  For a BH seed to form the
stellar cluster needs a total mass $\ge 260\,\msun$ in massive stars
that can segregate. We here assume that stars more massive than 10
$\msun$ efficiently segregate in the cluster core. Once all these more
massive stars are consumed collisions are assumed to stop. For our
chosen IMF a minimum VMS mass of $260\,\msun$, corresponds to a
minimum cluster mass of $1.5\times 10^4\,\msun$. We here include this
caveat and assume that BH seeds can form only in those nuclear
clusters more massive than this threshold.

The ability of runaway collisions to form an intermediate mass BH
also depends on the metallicity of the gas from which the stars
originated. Recent simulations of star-star collisions have shown that
at metallicities higher than $10^{-3}\, Z_{\odot}$, physical
collisions are too disruptive, and the build up of a star with mass
greater than 260 $\msun$ is inhibited (Glebbeek et al. 2009). For this
reason we assume that a BH seed forms only at metallicities 
$<10^{-3}\,Z_{\odot}$.

Summarising, BHs form provided that the following conditions are met:
(i) $Z>Z_{\rm crit}$ so that a NC can form; (ii) $\lambda$ is
sufficiently small for the disc to reach gravitational instability;
(iii) clusters form with $t_{\rm cc}< 3$ Myr and (iv) massive enough
to host an adequate amount of stars with $m_*>10\,\msun$.

\begin{figure}\label{fig:time}
\includegraphics[width=0.48\textwidth]{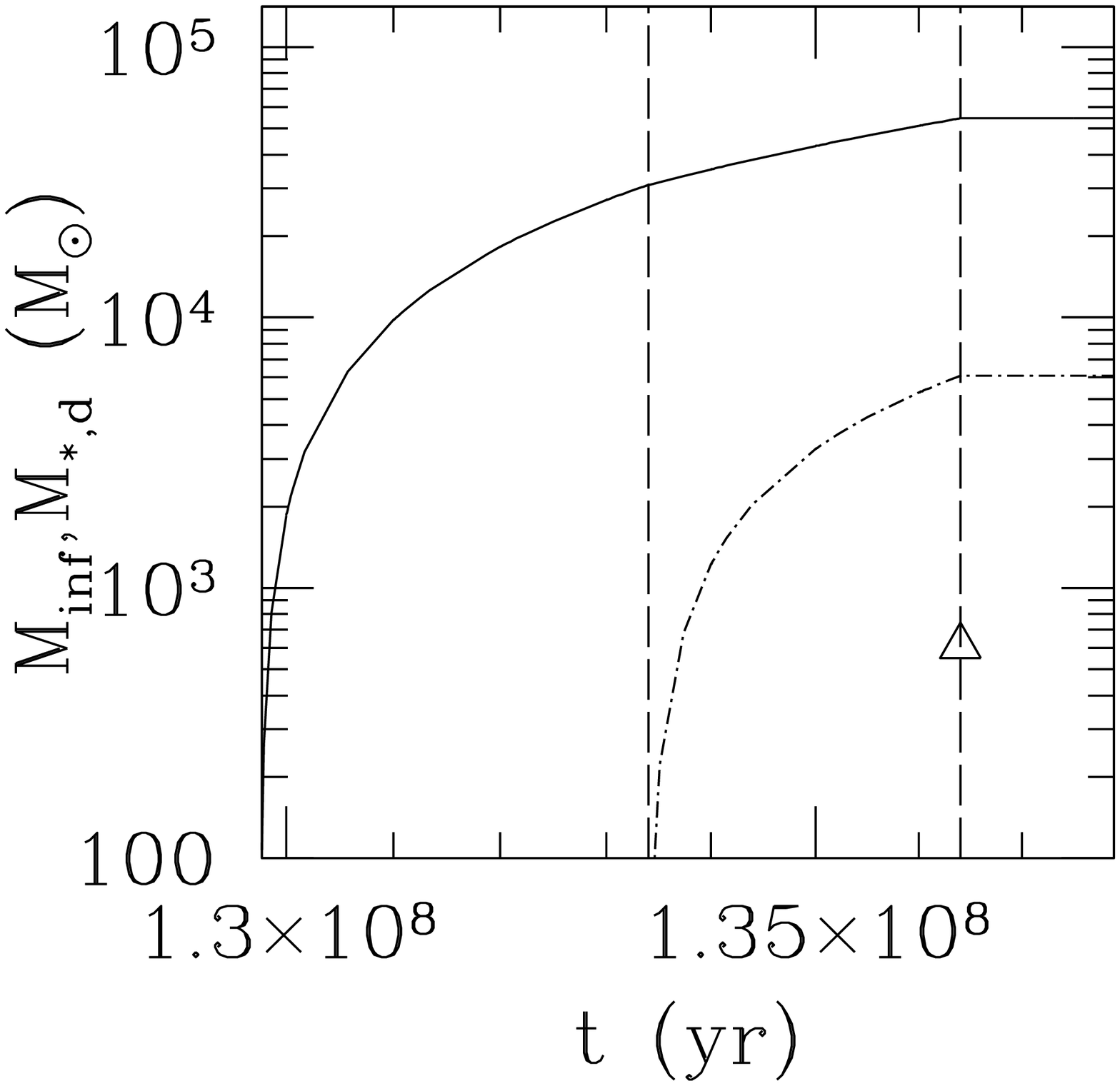}
\includegraphics[width=0.48\textwidth]{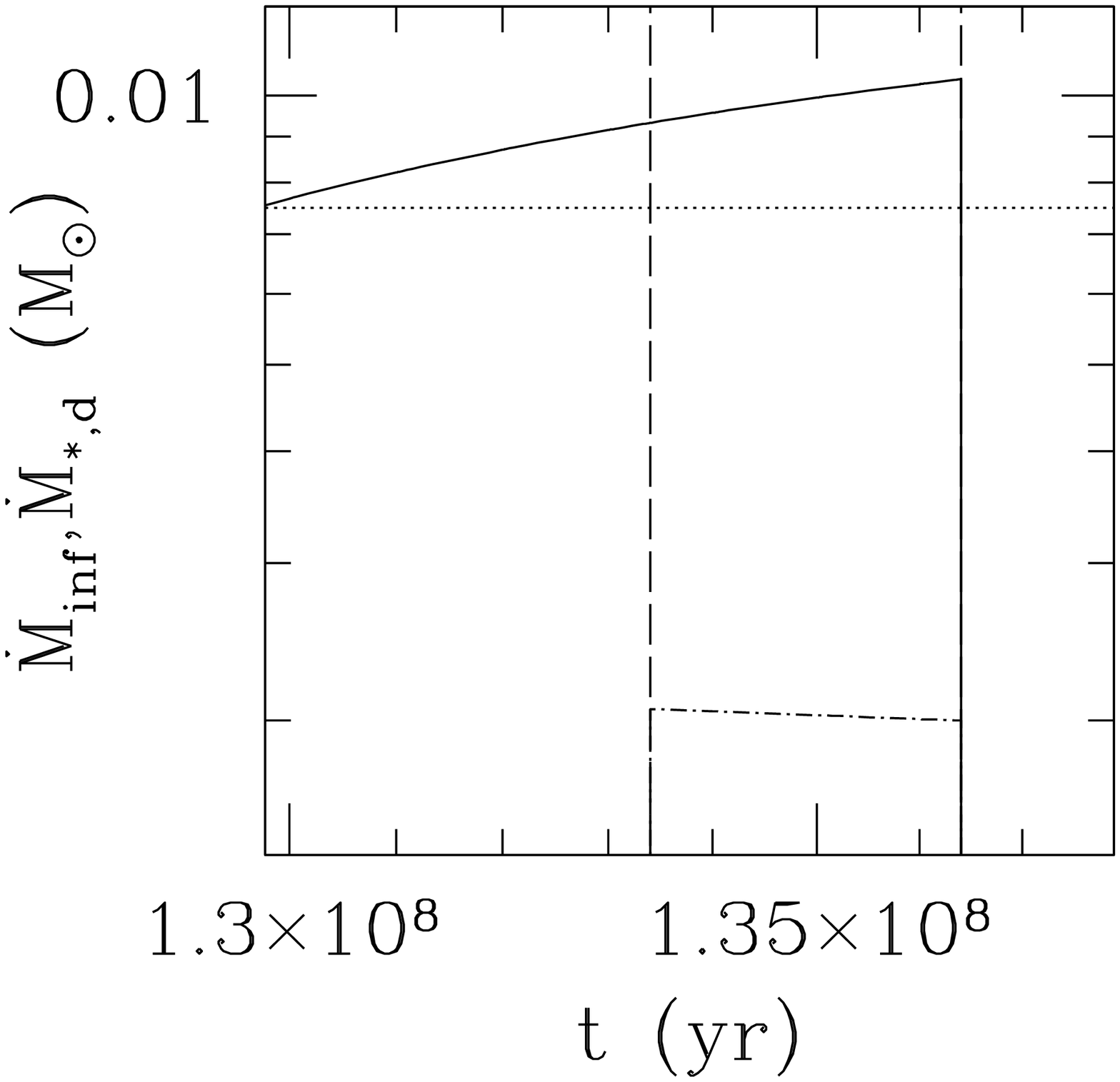}
\caption{Upper panel: $M_{\rm inf}$ (solid line) and $M_{\rm *,d}$
  (dash-dotted lines) versus time $t$ for our reference
  models. The triangle highlights the BH mass at the moment of its
  formation. Lower panel: $\dot{M}_{\rm inf}$ (solid line) and
  $\dot{M}_{\rm *,d}$ (dash-dotted line) for our reference model as a
  function of time. The horizontal dotted line marks the critical
  threshold for fragmentation $\dot{M}_{\rm crit}$.  Vertical lines
  define the different evolutionary phases of the disc properties as
  described in the text. }
\end{figure}

\section{Results}

In this Section we describe how disc properties are shaped
by gravitational instabilities and star formation.  Our reference
halo has mass $M_\h=6.5\times 10^7\,\msun$,
and forms at $z_{\rm vir}=10:$ we further adopt 
$\lambda=0.05$, $Z/Z_{\odot}= 10^{-4}$,  $\eta=0.3$,
$Q_{\rm c}=2$, $\epsilon_{\rm SF}=1$ and $t_{\rm lag}=3.5$ Myr.

\subsection{Disc and cluster properties}

In this subsection we illustrate the evolution of the reference disc
model. After a time $\tau_{\rm inst}=1.3\times 10^8$ yr, the disc has grown
enought to hit the instability threshold as it has assembled enough
mass so that the Toomre parameter drops below $Q_{\rm c}$.  The upper
(lower) panel in Fig. 1 shows $M_{\rm inf}$ ($\dot{M}_{\rm inf}$) and
$M_{\rm *,d}$ ($\dot{M}_{*,\rm d}$ ) as a function of time $t$, from
the time the disc becomes unstable until the time of formation of the
central BH. The evolution proceeds along different phases:

\begin{itemize}
\item[1)] $\tau_{\rm inst}<t<\tau_{\rm inst}+t_{\rm lag}$: at
  $t=\tau_{\rm inst}$ the disc starts collecting gas into the center
  and this leads to an enhancement of $M_{\rm inf}$. The disc
  parameters are such that $\dot{M}_{\rm grav}$ is greater than
  $\dot{M}_{\rm crit}$ (horizontal dotted line in the righ panel of
  the Figure). Clump formation in the disc sets in but these newly
  formed clouds have not had enough time to collapse until a time
  $t_{\rm lag}$ has elapsed.

\item[2)]$\tau_{\rm inst}+t_{\rm lag}<t<\tau_{\rm inst} + t_{\rm
  lag}+3\, {\rm Myr}$: star formation begins. The critical density for star
  formation remains constant as metals have not been released yet and
  the star formation radius still remains close to its initial
  value. Both $M_{\rm inf}$ and $M_{\rm *,d}$ increase until the first
  SNae explode. We label the nuclear mass at this time as $M_{\rm
    cl,0}$.

\item[3)]$t>\tau_{\rm inst} + t_{\rm lag}+3\, {\rm Myr}$: the first
  SNae explode affecting the subsequent evolution of the system. Part
  of the metals produced in stars are released in the surrounding gas
  increasing the metallicity above $Z_{\rm crit}$.  
  At the same time SN explosions evacuate gas from
  the system lowering the surface density. As a
  result the Toomre parameters increases above $Q_{\rm c}$.
 \end{itemize}
 
The disc then lingers on a marginally stable configuration thanks to
the high $\dot{M}_{\rm sh}$ that is able to counter-balance
$\dot{M}_{\rm cool}$. Fragmentation in the disc is temporary halted as
$\dot{M}_{\rm grav}$ drops below $\dot{M}_{\rm crit}$.  At later
times, the surface density starts to increase again, as hot gas in the
halo continues to cool.  The disc then enter a new cycle of
instability, and evolution terminates when the hot gas reservoir has
been consumed.

\subsection{Dependence on input parameters}

\begin{figure}\label{fig:mvs}
\includegraphics[width=0.4\textwidth]{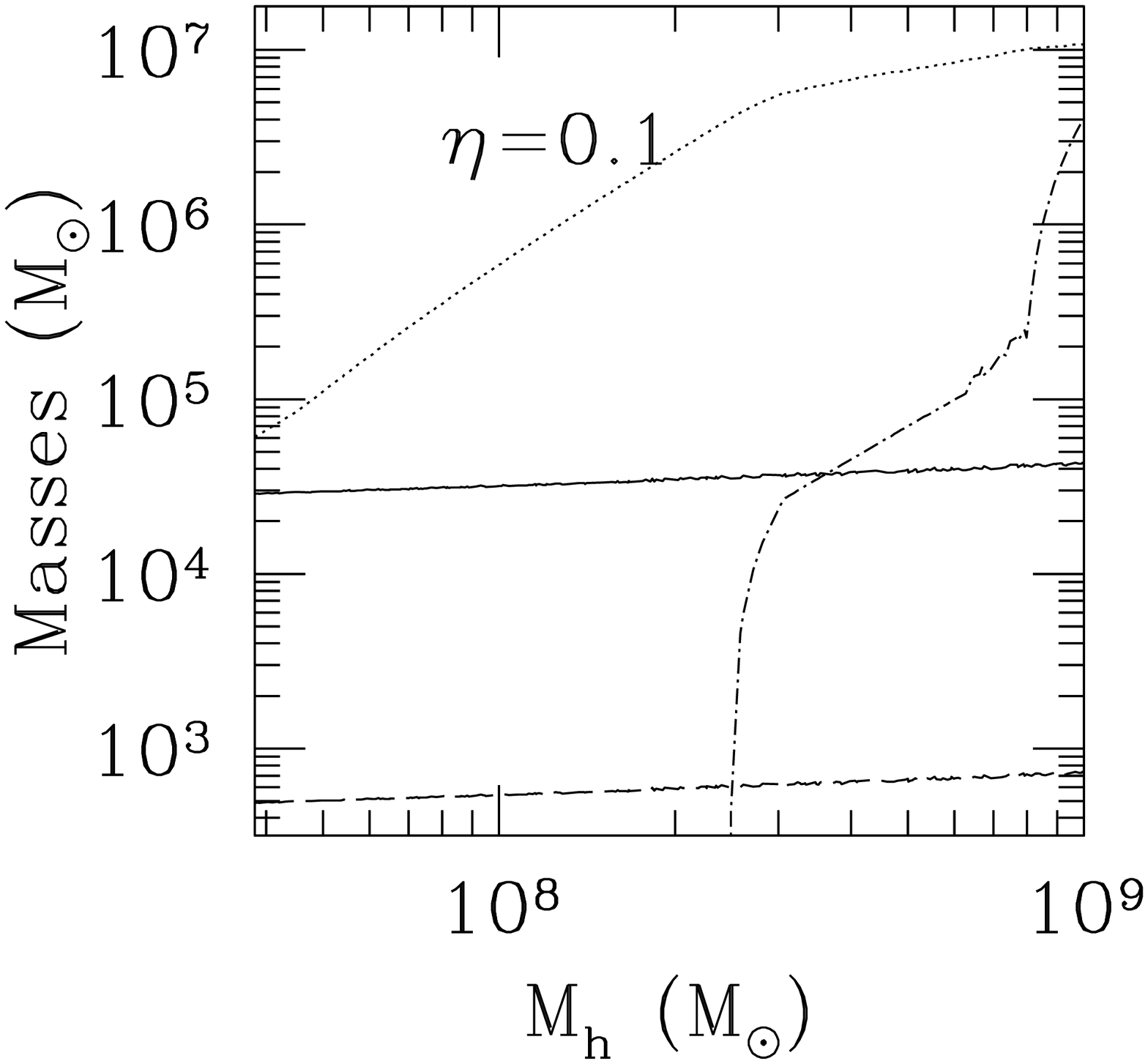}
\includegraphics[width=0.4\textwidth]{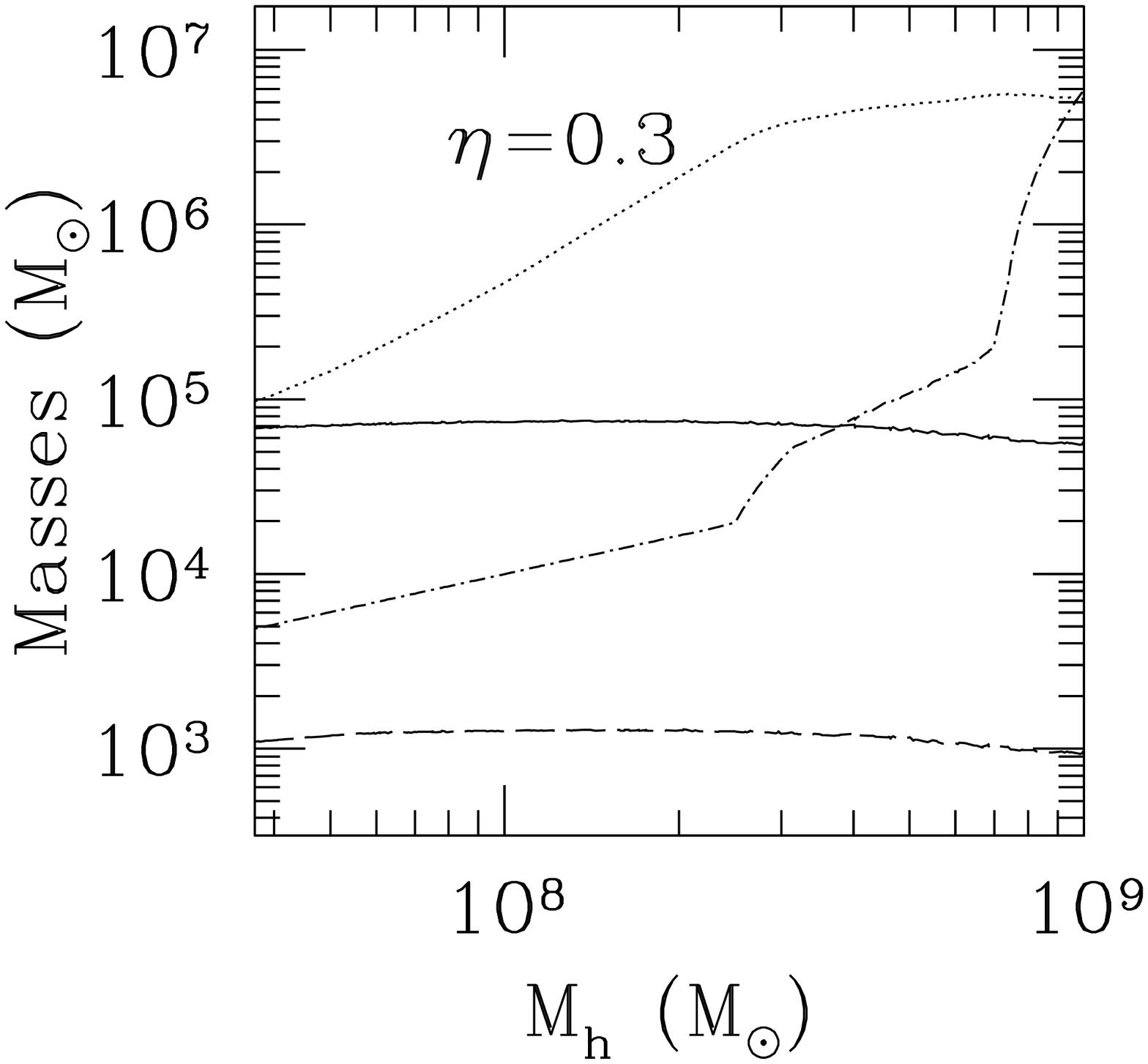}
\includegraphics[width=0.4\textwidth]{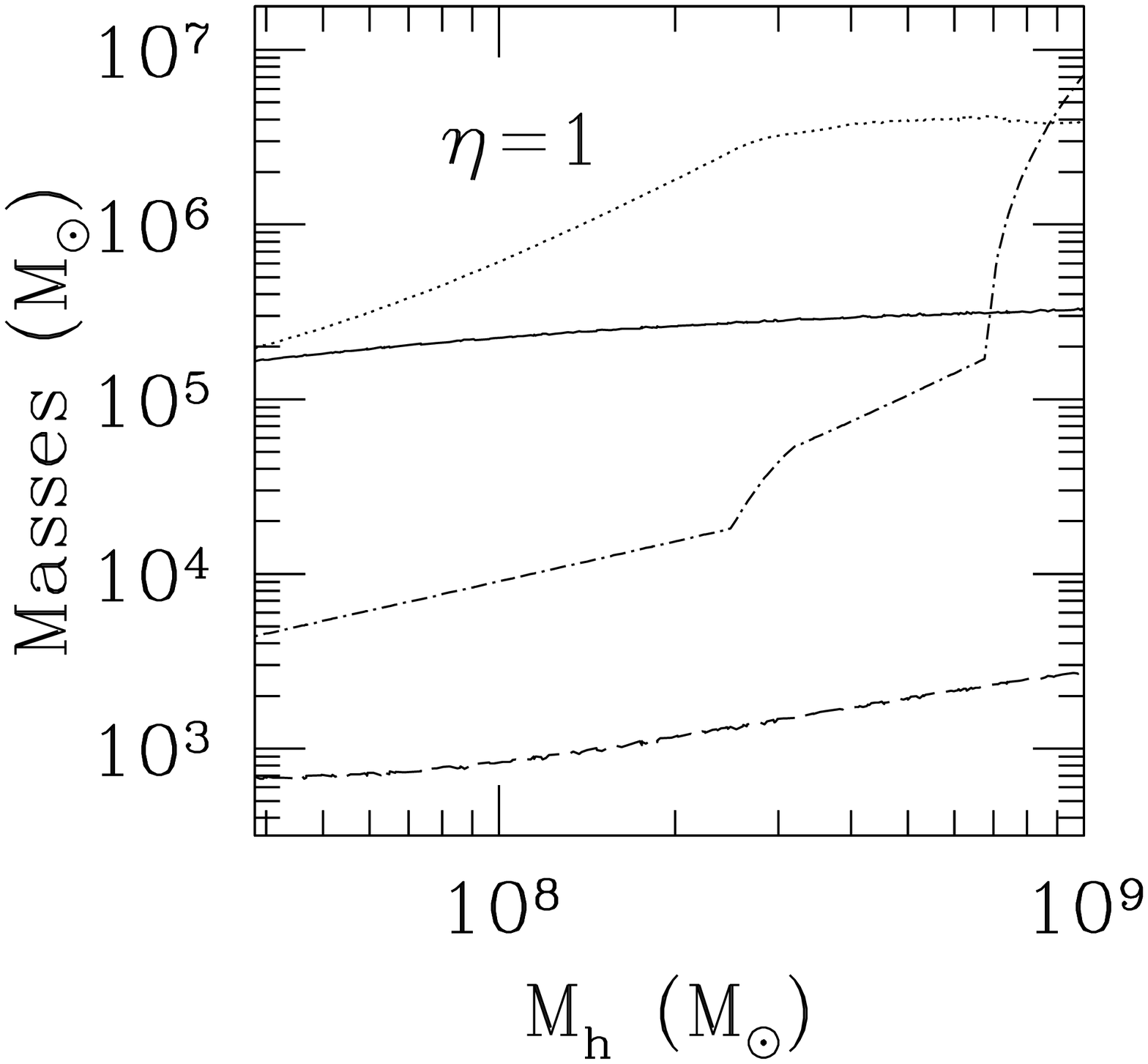}
\caption{Mass $M_{\rm cl,0}$ (solid lines) of the NC and $M_{\rm
    BH}$ (dashed lines) at the time the BH forms, and 
  $M_{\rm inf}$ (dotted lines), $M_{*,\rm d}$ (dash-dotted lines) at
  the end of the simulations, as a function of $M_\h$ for different
  inflow efficiencies.  High, central and low panels refer to
  $\eta=0.1$, 0.3 and 1, respectively. Reference properties for the
  halo are $z_{\rm vir}=10$ and $\lambda=0.05$; for the baryonic
  component $Z/Z_{\odot}= 10^{-4}$, $Q_{\rm c}=2$ and $\epsilon_{\rm
    SF}=1$.}
\end{figure}
We here explore disc evolution as a function of the halo mass, $M_\h$,
for three different values of inflow efficiency, $\eta$, and as a
function of the initial metallicity $Z$.

Fig. 2 shows the NC mass $M_{\rm cl,0}$ (solid lines) and BH mass
$M_{\rm BH}$ (log dashed lines) at the time of BH formation, and the
stellar disc mass $M_{*,\rm d}$ (dash-dotted lines), and net inflow
mass $M_{\rm inf}$ (dotted lines) at the end of the simulation, as a
function of $M_\h$, for different values of $\eta$.
 
The NC mass, $M_{\rm cl,0}$, depends weakly on the host mass, $M_\h$,
and somewhat more strongly on the inflow efficiency, $\eta$.  By
contrast, the final stellar disc mass depends significantly on $M_\h$.
This is because (i) more massive halos have a larger gas mass and
retention efficiency ($f_{\rm ret}$); (ii) gas cools more rapidly in
larger halos.  This allows for a larger departure of the Toomre
parameter $Q$ from $Q_{\rm c}$, and larger inflow ($\dot{M}_{\rm
  grav}$, see equation \ref{eq:inflow}). $\dot{M}_{\rm grav}$ is
larger than the critical inflow triggering fragmentation,
$\dot{M}_{\rm crit}$, and more stars form in the outer disc.  For
$\eta=0.1$ a threshold in $M_\h$ exists below which $\dot{M}_{\rm
  grav}<\dot{M}_{\rm crit}$ always, and $M_{\rm *,d}$ drops to zero.

Dotted lines in Fig. 2 show the net inflow mass, $M_{\rm inf}$, at the
end of the simulations. $M_{\rm inf}$ increases with increasing $M_\h$
as a result of both initially larger gas masses and deeper potential
wells. $M_{\rm inf}$ depends weakly on $\eta$. For higher $\eta$ more
gas is consumed by extended star formation before it can reach the
nucleus and final $M_{\rm inf}$ are lower.

\begin{figure}\label{fig:mvs2}
\includegraphics[width=6cm]{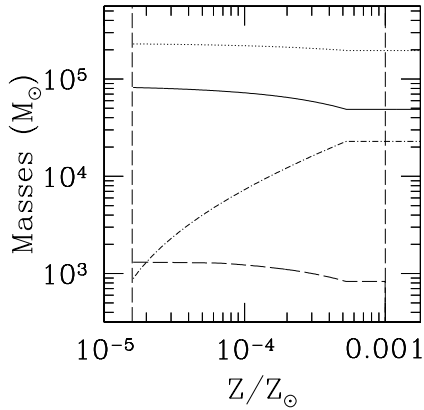}
\caption{$M_{\rm cl,0}$ (solid lines), $M_{\rm BH}$ (dashed lines),
  $M_{\rm inf}$ (dotted lines) and $M_{*,\rm d}$ (dash-dotted lines)
  as a function of the initial metallicity of the gas for our
  reference halo. The halo mass is kept constant at its reference
  value $M_\h=6.7\times 10^7\,\msun$. Vertical lines mark the minimum
  metallicity for low-mass star formation, $Z_{\rm crit}\simeq
  10^{-4.87}\,Z_{\odot}$ , and the maximum metallicity that allows BH
  formation, $Z_{\rm crit}\simeq 10^{-3}\,Z_{\odot}$.}
\end{figure}

Disc evolution depends also on the initial metallicity $Z.$ The
metallicity determines when stars can form and it is thus a key
parameter: there is a minimum metallicity where `normal' stars star
forming, $Z_{\rm crit}\simeq 10^{-4.87}\,Z_{\odot}$, and a maximum
metallicity that allows BH formation, $Z_{\rm crit}\simeq
10^{-3}\,Z_{\odot}$.

In Fig.  3, the NC mass $M_{\rm cl,0}$ (solid line), BH mass $M_{\rm
  BH}$ (dashed line), net inflow mass $M_{\rm inf}$ (dotted line), and
stellar disc mass $M_{*,\rm d}$ (dash-dotted line) are plotted as a
function of the initial gas metallicity. Star formation sets in only
for $Z>Z_{\rm crit}$ (left vertical dashed line in Fig. 3). For the
majority of the systems we are considering here, $R_{\rm SF}>R_{\rm
  tr}$.  The first three quantities are almost independent of
metallicity, as long as $Z$ is in the range between the minimum and
maximum value. The disc stellar mass is instead very sensitive to the
exact value of the metallicity. $M_{\rm *,d}$ increases with $Z$ (see
equation~\ref{eq:sfrate}). Mass is consumed instead of flowing into
the nucleus, and $M_{\rm cl,0}$ weakly decreases\footnote{Note that
  for systems with $R_{\rm SF}<R_{\rm tr}$, the initial cluster mass,
  $M_{\rm cl,0}$, increases with $Z$ as (i) extended star formation is
  not allowed in the outer disc, so that the amount of mass funnelled
  into the nucleus remains constant; (ii) higher metallicities
  correspond to higher $R_{\rm SF}$ so that to a larger amount of
  nuclear gas can be converted into stars. This case was considered in
  DV09, we refer to that work for a discussion of this regime.}.  When
the metallicity is large enough that $R_{\rm SF}$ equals $R_0$ than
the whole disc forms stars, and the dependence on $Z$ disappears.

In our reference model we set $\lambda=0.05$. The distribution of spin
parameters can be described as a log-normal distribution with peak at
$\lambda\sim 0.03-0.05$ and dispersion $\sim 0.5$. For our reference
halo the maximum $\lambda$ allowing for instability is 0.07.
Approximately 96\% of the halos in the interesting mass range have
$\lambda<\lambda_{\rm max}$ and are therefore prone to disc
instabilities.

If $Q_{\rm c}$ is lowered to unity, instabilities are easily quenched,
and the process becomes less efficient (see upper right panel in
Fig. 4).

\subsection{NC and BH masses versus redshift}

\begin{figure*}\label{fig:confronto}
\begin{center}
\includegraphics[width=6cm]{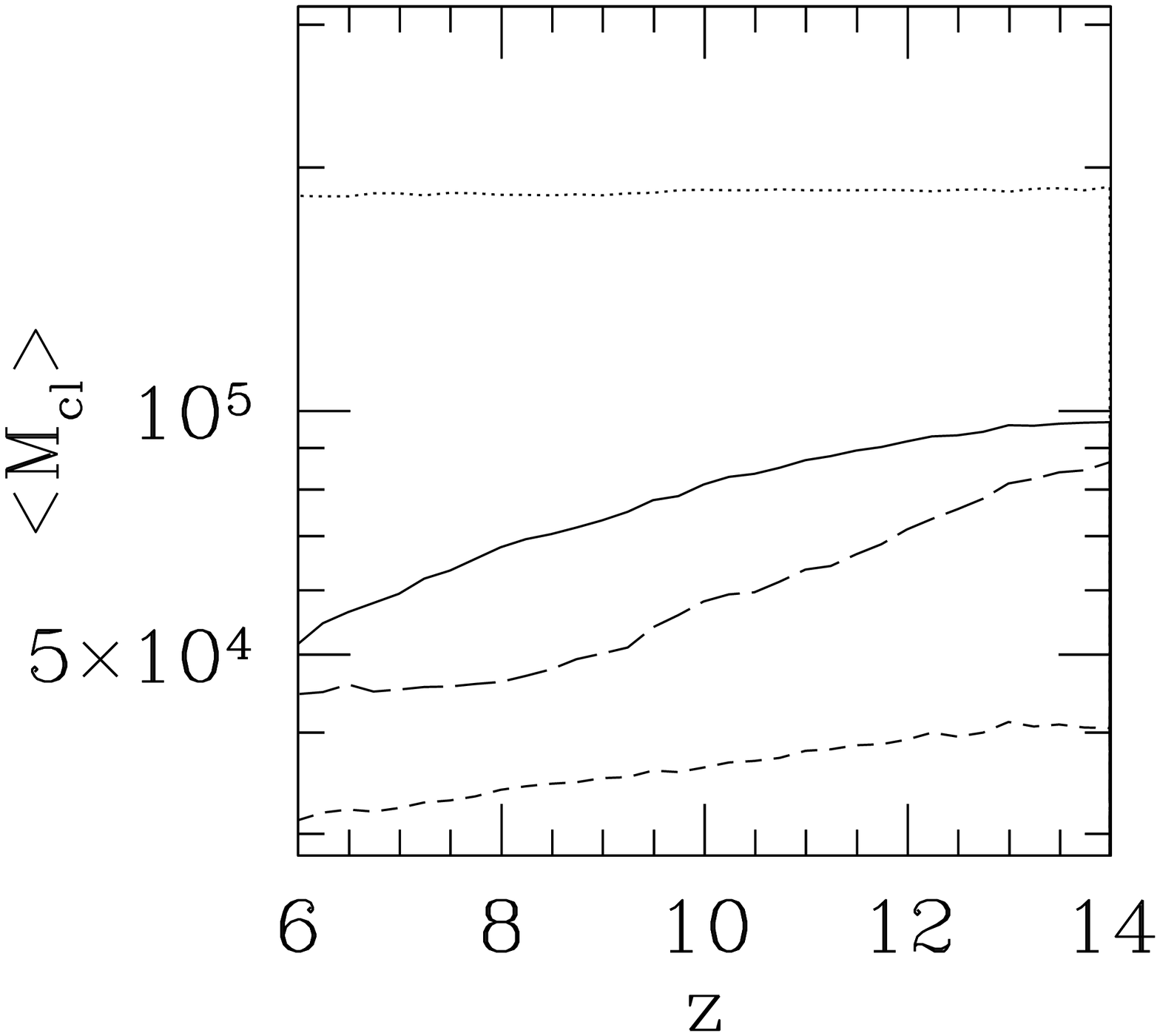}
\includegraphics[width=6cm]{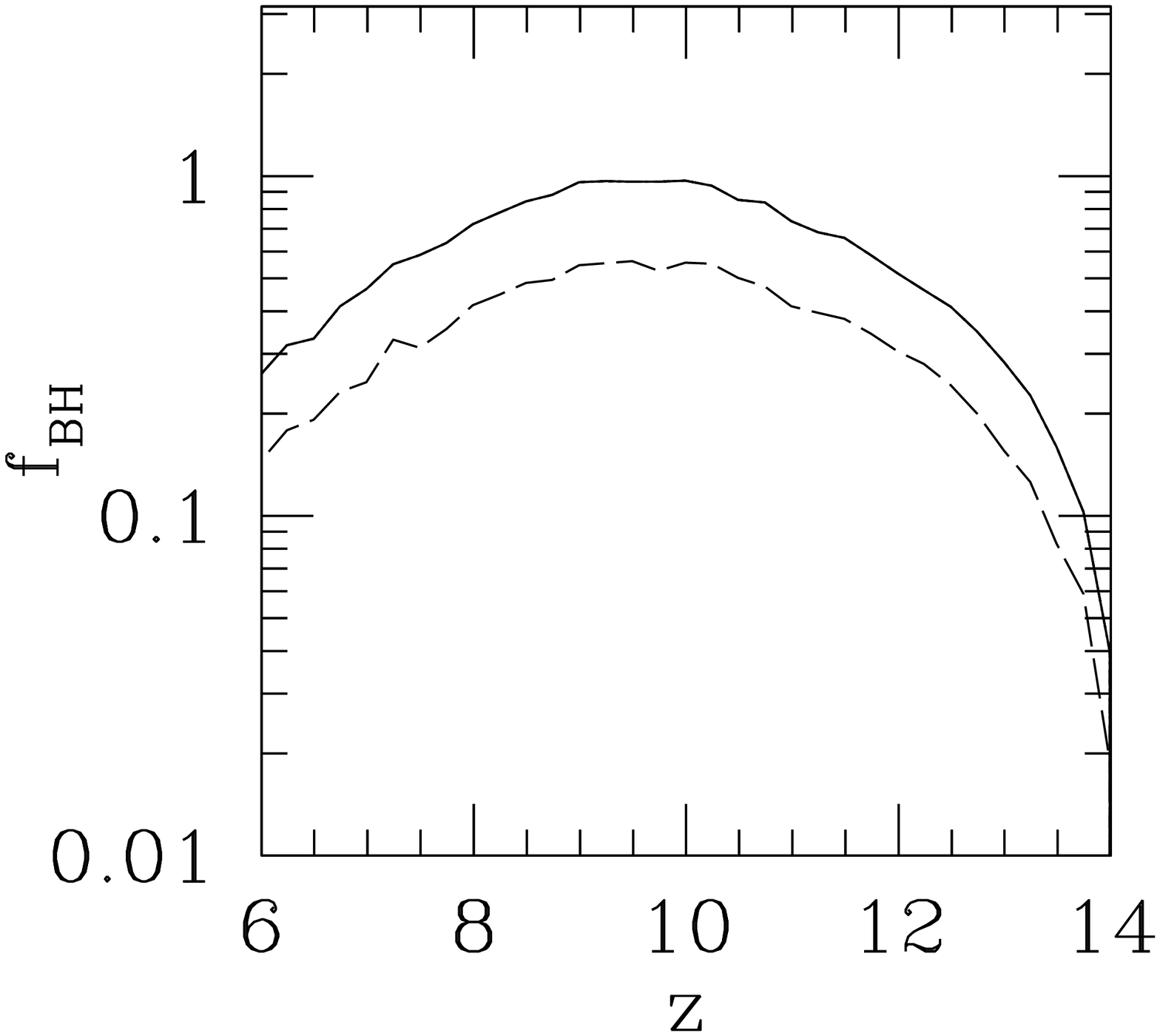}
\includegraphics[width=6cm]{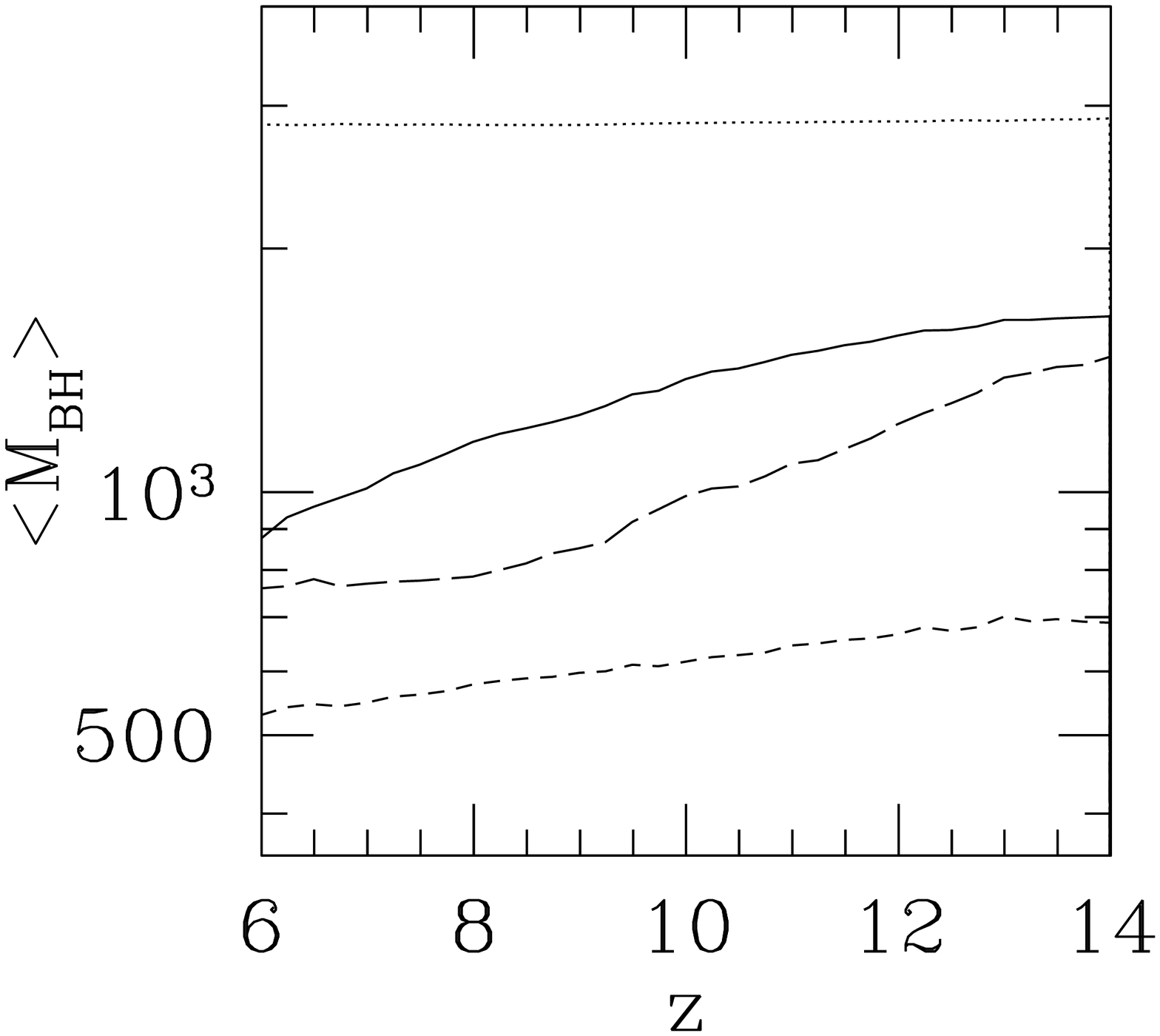}
\includegraphics[width=6cm]{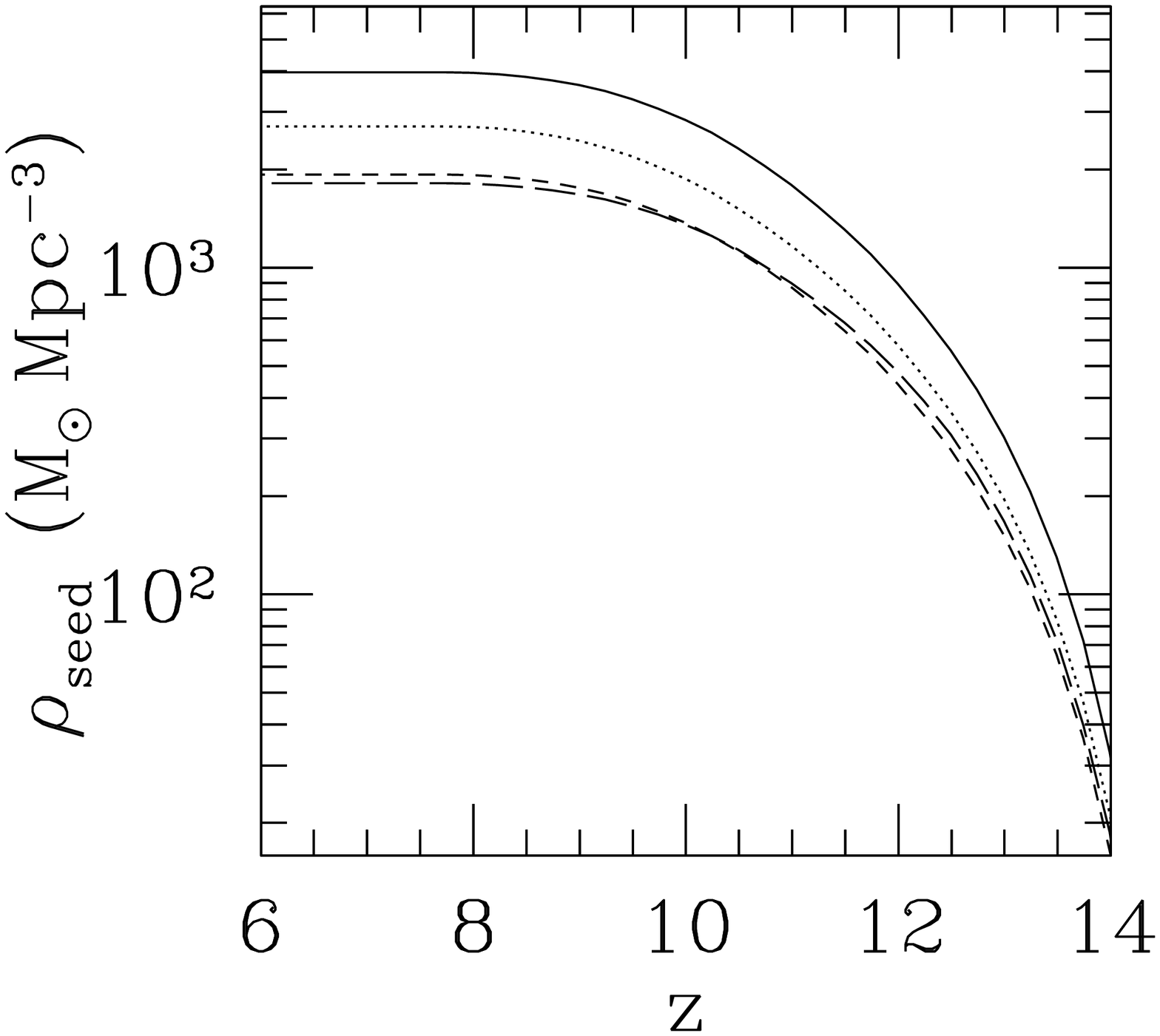}
\end{center}
\caption{ Panels from up to low: $\langle M_{\rm cl,0} \rangle$ (for
  those clusters able to form a seed BH), $f_{\rm BH}$, $\langle
  M_{\rm BH}\rangle$ and $\rho_{\rm seed}$ as a function of redshift
  for different values of $\eta$ and $Q_{\rm c}$. Line styles: solid
  lines refer to a model with $\eta=0.3$, $Q_{\rm c}=2$, dotted lines
  refer to a model with $\eta=1$, $Q_{\rm c}=2$, short-dashed lines
  refer to a model with $\eta=0.1$, $Q_{\rm c}=2$, long-dashed lines
  refer to a model with $\eta=0.3$, $Q_{\rm c}=1$.}
\end{figure*}

In this Section we apply our prescriptions for the cooling of gas,
star formation and inflow to determine the properties of the NC and BH
populations, in low metallicity environments.

To this purpose, we construct a sample of halos with the same
properties of those analysed in DV09. We focus on halos with $z_{\rm
  vir}$ ranging between 5 and 30 corresponding to $T_{\rm vir}$
between $10^4$ K and $1.8\,\times\,10^4$ K (Barkana \& Loeb 2001), and
we determine their frequency using a modified version of the Press \&
Schechter formalism (Sheth \& Tormen 1999) in a WMAP5 cosmology
(Dunkley et al. 2009). To each halo, we assign a value of the spin
parameter, $\lambda$, extracted from the probability distribution
found in the Millennium simulations. We here adopt the same
prescriptions for the metallicity evolution as in model A in DV09,
while we adopt the model described in this paper for the evolution of
each disc-halo. The model is thus an improvement of that discussed in
DV09. More realistic models that follow metal enrichment
self-consistently will be described in Paper II.  We consider
different inflow efficiencies $\eta$=1 (dotted lines), 0.3 (solid
lines) and 0.1 (short-dashed lines) for $Q_{\rm c}=2$ and $\eta=0.3$
for $Q_{\rm c}$=1 (long-dashed lines).

In the upper left panel of Fig. 4 we show the evolution of the NC
mass, $\langle M_{\rm cl,0} \rangle$ (averaged over the sample) as a
function of redshift. NCs start to form at $z=14$, as halos are
polluted above the critical metallicity for fragmentation 
(see DV09)\footnote{In our reference model ($\eta=0.3$, $Q_{\rm c}=2$) the mean
  cluster mass $\langle M_{\rm cl,0} \rangle$ decreases with redshift
  as the metallicity increases. This behaviour appears only as the
  majority of the systems we are considering here have $R_{\rm
    SF}>R_{\rm tr}$ (that is, stars form in most of the disc).  If
  this were not the case, the behaviour of $\langle M_{\rm
    cl,0}\rangle$ versus redshift would be the opposite (see DV09).}.
BH seeds start to form after the formation of the first NCs. In the
upper right panel of Fig. 4 we compute the fraction $f_{\rm BH}$ of
halos with $T_{\rm vir}>10^4$ K hosting a BH seed, as a function of
redshift. $f_{\rm BH}$ is regulated by the fraction of halos with
$Z_{\rm crit}<Z<10^{-3}\,Z_{\odot}$: it first increases as more and
more halos are polluted, and then decreases when most halos reach
$Z=10^{-3}Z_{\odot}$.  $f_{\rm BH}$ in this paper are larger than
those found in DV09 (see their Figure 8, upper panel) by a factor 10.
With the prescriptions we adopt here, $M_{\rm cool}$ evolves with
time, regulated by cooling of the halo gas whereas in DV09 a fixed
$M_{\rm cool}/M_\h$ is used. This implies that for a given $Q_{\rm
  c}$, the fraction of unstable discs that form is increased.

Mean black hole masses $\langle M_{\rm BH}\rangle$ versus redshift $z$
are plotted in Fig. 4 (lower left panel). $\langle M_{\rm BH}\rangle$
spans values between a few hundred $\msun$ and 2000 $\msun$.  For most
of the NCs considered here, the growth of the VMSs is limited by the
amount of stars with $m_*>10\,\msun$. Lower $\langle M_{\rm
  cl,0}\rangle$ thus lead to lower mass seed BHs. An exception occurs
for the highest $M_{\rm cl,0}$ (i.e. $\eta=1$): for these systems
$t_{\rm cc}$ is long and not all massive stars contribute in building
up the VMS. $\langle M_{\rm BH}\rangle$ stays at a near constant value
$\sim$ 900 $\msun$, at any redshift.

Comoving BH seed densities are plotted in the lower right panel.
$\rho_{\rm seed}$ depends on both BH formation efficiencies and masses
attained. They are found to depend weakly on the parameters: final
$\rho_{\rm seed}$ only differs of a factor 2, ranging between 2000 and
4000 $\msun\,{\rm Mpc}^{-3}$.

\section{Summary}

In this paper, the first of a series, we presented a model for the
formation of high redshift, metal poor NCs and their embedded BH
seeds. The channel that we propose can be summarized as follows:

\begin{itemize}
\item After the virialisation of the halo, hot gas cools down and the
  presence of angular momentum leads to the formation of a cold
  gaseous disc. The disc accretes gas from the halo and grows in mass,
  increasing its surface density to the extent that gravitational
  instabilities develop. This leads to the growth of non-axisymmetric
  perturbations that redistribute angular momentum, driving mass into
  the central part of the halo.
\item For those systems where the metallicity $Z$ exceedes $Z_{\rm
  crit}$ and $\dot{M}_{\rm grav}>\dot{M}_{\rm crit}$, star formation
  in the extended disc can set in, consuming part of the mass that
  would otherwise been accreted in the center. The inner density
  profile steepens inside a transition radius where a compact nuclear
  stellar cluster forms.
\item In those NCs where $t_{\rm cc}<3$ Myr, core collapse anticipates
  stellar evolution and runaway collisions between the most massive
  stars build up a very massive star that finally implodes leaving a
  remnant BH seed.
\item The evolution continues and it is driven by the competition
  between star formation, triggered by prolonged cooling
  of hot halo gas and SN explosions that evacuate part of it. 
\item Evolution halts as soon as the hot halo gas is consumed/blown
  away and the disc settles into a state of stationary equilibrium.
\end{itemize}

We apply our recipes to a set of halos with $T_{\rm vir}>10^4$ and
properties typical of $z\sim$10-20.  We determine how changing halo
properties and model parameters influence the final NC and BH seed
population. The NCs have masses, at the moment of BH formation, of
$10^4-10^6\,\msun$ and host BHs with masses between 300 and a few
$10^3\,\msun$.

We compared our new prescriptions with the results of DV09. The mean
NC masses are found to decrease with decreasing redshift: higher
metallicities usually correspond to larger star formation rates in
the extended disc. These reduce the amount of gas channelled into the
center, leading to less massive clusters. Except for the more massive
NCs, $M_{\rm BH}$ is dictated by the availability of massive
stars. The decrease in $M_{\rm cl,0}$ consequently leads to less
massive black holes.  The total mass densities in BH seeds, $\rho_{\rm
  seed}$, are higher with respect to DV09 as a result of the larger
number of gravitationally unstable discs. In our model we do not take
into account any BH growth by accretion, nor their assembly during
galaxy clustering. Comoving densities found here are thus lower
limits.

In the second paper of this series we will apply this semi-analytical
scheme to merger tree histories to trace NC and BH seed formation into
a cosmological context. In the third, we will follow the evolution of
the emerging BH population through accretion and clustering and
compare our results with current observation of BH occupation
frequency in low redshift galaxies, quasar luminosity functions, etc.

\appendix

\section{Gas depletion from SN feedback}

In this section we describe the prescription used to infer the amount
of mass lost from the host following a SN episode.  Under the
assumption that the gas follows an isothermal profile

\begin{equation}\label{eq:rho}
\rho_{\rm gas}=\frac{\Omega_{\rm b}}{\Omega_{\rm m}}\frac{V^2_\h}{4\pi
  Gr^2},
\end{equation}
we compute the initial binding energy of the system $E_{\rm b,in}$
inclusive of the dark matter and gaseous component\footnote{Note that
  the actual gas density profile at the moment of SN explosions would
  be quite different from the initial one. Differences would arise
  both from the fact that the gas needs to cool down in order to form
  stars, and from the radiative feedback of stars during their
  lifetimes.}.

\begin{equation}\label{eq:ebinin}
E_{\rm b,in}=-\frac{GM^2_\h}{R_{\rm vir}}\left(1+\frac{\Omega_{\rm
    b}}{\Omega_{\rm m}}\right)^2.
\end{equation}

After removing a gaseous mass $M_{\rm sh}$ the new binding $E_{\rm
  b,f}$ is

\begin{equation}\label{eq:ebinf}
E_{\rm b,f}=-\frac{GM^2_\h}{R_{\rm vir}}\left(1+\frac{\Omega_{\rm
    b}}{\Omega_{\rm m}}-\frac{M_{\rm sh}}{M_\h}\right)^2.
\end{equation}

Energy conservation imposes the following equality: $E_{\rm
  b,in}+\mathcal{E}_{\rm SN}=E_{\rm b,f}+K_{\rm SN}$, where
$\mathcal{E}_{\rm SN}$ is the energy injected by the SN explosions and
$K_{\rm SN}$ is the kinetic energy of the outflow that is leaving the
halo .

The energy $\mathcal{E}_{\rm SN}$ by SNae, over a time $t,$ is given
by

\begin{equation}\label{eq:esnpopii}
\mathcal{E}_{\rm SN}(t)=\nu_{\rm SN} E_{\rm SN}\int^t_0 \dot{M}_{*}{\rm d}s
\end{equation}
\noindent
where $E_{\rm SN}$ is the energy of a single SN explosion, set equal
to $10^{51}$ erg, $\dot{M}_{\rm *}=\epsilon_{\rm SF}\dot{M}_{\rm
  inf}+\dot{M}_{*,\rm d}$ and $\nu_{\rm SN}$ is the ratio between the
total number of SNae exploding after the formation of a mass $M_*$ of
stars, divided by $M_*$. SNae start to explode after 3 Myr from the
onset of the star formation episode. We assume that at that time all
stars with masses in the SN regime explode together giving rise to a
single bubble and that a fraction $f_w$ of the total energy released
by SN explosions, is channeled into the outflow.  Following
Scannapieco et al. (2003), we relate $f_w$ with the
halo mass as follows. We define $\delta_{\rm B}(M_\h)$ as

\begin{equation}
  \delta_{\rm B}(M)= \left\{
\begin{array}{lr}
1.0, & \tilde{N}_{\rm t}\leq 1\\ 1.0-0.165\ln (\tilde{N}^{-1}_{\rm
  t}), &1\leq \tilde{N}_{\rm t}\leq 100 \\ \left[ 1.0-0.165\ln
  (100)\right] 100\tilde{N}^{-1}_{\rm t} & 100\leq\tilde{N}_{\rm t}\\
\end{array}
\right.
\label{eq:deltaB}
\end{equation}
\noindent
where $\tilde{N}_{\rm t}\equiv 1.7\times 10^{-7} (\Omega_{\rm
  b}/\Omega_{\rm m}) M_\h/\msun$. $f_w$ is then given by
$f_w=\delta_{\rm B}(M_\h)/\delta_{\rm B}(M_\h=2\times 10^8
\msun)$. This prescription follows from the scaling relation found in
Ferrara et al. (2000), normalised with the result found by Mori et
al. (2002).

Let us now compute $K_{\rm SN}\equiv 1/2 M_{\rm sh}v^2_{\rm sh}$. Here
the shell velocity is calculated at the moment it reaches $R_{\rm
  vir}$.  The topology of early metal enrichment can be extremely
complex.  Enriched gas first propagates into the cavities created by
the shock (Greif et al. 2007), and after the bubble shell leaves the
host, it spreads mainly into voids, leading to a very anisotropic
spatial distribution of metals. We here neglect these degrees of
complexity and simply assume that SN-driven bubbles evolve in
spherical symmetry. The shell radius is assumed to
follow the Sedov-Taylor solution, as long as the pressure of the
bubble drops below the value of the surrounding medium.
The shell radius $R_{\rm sh}$ evolves with time $t$ as

\begin{equation}
R_{\rm sh}=1.15\left(\frac{f_w \mathcal{E}_{\rm SN}t^2}{\rho_{\rm
    bk}}\right)^{1/5}
\label{eq:rsn}
\end{equation}
\noindent
where $t$ is the time after the explosion and $\rho_{\rm bk}$ is the
ambient gas density, scaling as $(\Omega_{\rm b}/\Omega_{\rm
  m})M_\h/R^3_{\rm vir}$. The velocity of the shell $v_{\rm sh}$ in
the Sedov-Taylor approximation is given by $2/5 R_{\rm sh}/t$. Under
these conditions, $v_{\rm sh}(R_{\rm vir})$ can be easily inferred.
At the time $R_{\rm sh}=R_{\rm vir}$ the kinetic energy is

\begin{equation}\label{eq:ksn}
K_{\rm SN}=0.67\frac{\Omega_{\rm m}}{\Omega_{\rm b}}\frac{M_{\rm
    sh}}{M_\h}f_w\mathcal{E}_{\rm SN}.
\end{equation}

To compute $M_{\rm sh}$ for a given halo we recure to energy
conservation leading to

\begin{equation}\label{implicita}
\left(1-0.67\frac{\Omega_{\rm m}}{\Omega_{\rm b}}\frac{M_{\rm
    sh}}{M_\h}\right)f_w\mathcal{E}_{\rm SN}=\frac{GM_\h M_{\rm
    sh}}{R_{\rm vir}}\left[2\left(1+\frac{\Omega_{\rm b}}{\Omega_{\rm
      m}}\right)-\frac{M_{\rm sh}}{M_\h}\right].
\end{equation}
\noindent
We note that $M_{\rm sh}/M_\h$ is always much less than $
2\left(1+\Omega_{\rm b}/\Omega_{\rm m}\right)$. Negletting the term
$M_{\rm sh}/M_\h$ in parentesis on the right side of the equation the
outflow mass of the first single bubble is given by

\begin{equation}\label{eq:mshpopiii}
M_{\rm sh}=\frac{f_w\mathcal{E}_{\rm SN}}{2\left(1+\Omega_{\rm
    b}/\Omega_{\rm m}\right)GM_\h/R_{\rm vir}+\frac{1}{2}v^2_{\rm
    sh}}.
\end{equation}
\noindent
We assume that an outflow develops only when $v_{\rm sh}>v_{\rm
  esc}$. Otherwise $M_{\rm sh}=0$.

Massive halos are not compleately evacuated after this first SN
episode. Further star formation is often allowed and to account for
subsequent SN esplosion episodes we compute the total number of
exploding SNae $N_{\rm SN}$ between time $t_1$ and $t_2$ 

\begin{equation}
N_{\rm SN}=\int^{t_2}_{t_1} \nu_{\rm SN} \left(\epsilon_{\rm
  SF}\dot{M}_{\rm inf}+\dot{M}_{*,\rm d}\right){\rm d}s.
\end{equation}

We define $M_{\rm sh,s}$ as the contribution of a single esplosion
(i.e. equation \ref{eq:mshpopiii} with $\mathcal{E}_{\rm SN}=E_{\rm
  SN}$).  An Outflow develops only if it injects sufficient energy so
that $v_{\rm sh}>v_{\rm esc}$. This requires a minimum energy in a
single bubble, i.e. a minimum number of SNae. For a given $\dot{M}_*$,
we calculate the expected number of SNae on a timescale of the order
of their progenitor lifetimes (i.e. 3 Myr). If the resulting $v_{\rm
  sh}$ is greater than the escape velocity from the halo, outflows are
allowed and
 
\begin{equation}
\dot{M}_{\rm sh}=M_{\rm sh,s}\frac{{\rm d}N_{\rm SN}}{{\rm
    d}t}=\frac{f_w\nu E_{\rm SN} \left(\epsilon_{\rm SF}\dot{M}_{\rm
    inf}+\dot{M}_{*,\rm d}\right)
}{2\left(1+\frac{\Omega_{\rm b}}{\Omega_{\rm m}}\right)\frac{GM_\h}{R_{\rm
      vir}}+\frac{1}{2}v^2_{\rm sh}}.
\end{equation}

Otherwise $\dot{M}_{\rm sh}=0$.


\end{document}